\documentstyle[preprint,aps,eqsecnum,psfig]{revtex}
\tightenlines
\def\m@thcombine#1#2{%
  \setbox0=\hbox{$#1$}
  \setbox1=\hbox{$#2$}
  \ifdim\wd0>\wd1
    \setbox0=\hbox to\wd1{\hss\box0\hss}
  \else
    \setbox1=\hbox to\wd0{\hss\box1\hss}
  \fi
  \mathop{\vcenter{
    \offinterlineskip\box0\box1}}}
\def\lesim{\m@thcombine<\sim}
\def\gesim{\m@thcombine>\sim}

\begin{document}

\draft
\title{ DETERMINATION OF THE QUANTUM PART OF THE TRULY NONPERTURBATIVE
YANG-MILLS VACUUM ENERGY DENSITY IN THE COVARIANT GAUGE QCD }

\author{V. Gogohia, Gy. Kluge }

\address{HAS, CRIP, RMKI, Theory Division, P.O.B. 49, H-1525 Budapest 114, Hungary \\
 email address: gogohia@rmki.kfki.hu}

\maketitle

\begin{abstract}
Using the effective potential approach for composite operators, we have formulated a general
method of calculation of the truly nonperturbative Yang-Mills vacuum energy density in the covariant gauge QCD ground state quantum models.
It is defined as an integration of the truly nonperturbative part of the full  
gluon propagator over the deep infrared region (soft momentum region).         
A nontrivial minimization procedure
makes it possible to determine the value of the soft cutoff in
terms of the corresponding nonperturbative scale parameter, which
is inevitably present in any nonperturbative model for the full
gluon propagator. We have shown for specific models of the full gluon
propagator explicitly that the use of the infrared enhanced
and finite gluon propagators lead to the vacuum energy density
which is finite, always negative and it has no imaginary part
(stable vacuum), while the infrared vanishing propagators lead to
unstable vacuum and therefore they are physically unacceptable.
\end{abstract}


\vfill
\eject

\section{Introduction}

The nonperturbative QCD vacuum is a very complicated medium and
its dynamical and topological complexity [1-3] means that its
structure can be organized at various levels (classical, quantum).
It can contain many different components and ingredients
which contribute to the truly nonperturbative vacuum energy
density (VED), one of the main characteristics of the QCD ground
state. Many models of the QCD vacuum involve some extra classical
color field configurations such as randomly oriented domains of
constant color magnetic fields, background gauge fields, averaged
over spin and color, stochastic colored background fields, etc.
(see Refs. [1,4,5] and references therein). The most elaborated
classical models are random and interacting instanton liquid
models (RILM and IILM, respectively) of the QCD vacuum [6]. These
models are based on the existence of the topologically nontrivial
instanton-type fluctuations of gluon fields, which are
nonperturbative, weak coupling solutions to the classical
equations of motion in Euclidean space (see Ref. [6] and
references therein).

Here we are going to discuss the quantum part of VED which is determined by the
effective potential approach for composite operators
[7-9]. It allows us to investigate the nonperturbative QCD
vacuum, in particular Yang-Mills (YM) one, by substituting some physically
well-justified Ansatz for the full gluon propagator since the exact
solutions are not known.
In the absence of external sources the effective potential is nothing
but VED which is given in the form of the loop expansion
where the number of the vacuum loops (consisting in general of the confining
quarks and nonperturbative gluons properly regularized with the help of ghosts
) is equal to the power of the Plank constant, $\hbar$.

The full dynamical information of
any quantum gauge field theory such as QCD is contained in the
corresponding quantum equations of motion, the so-called
Schwinger-Dyson (SD) equations for lower (propagators) and higher
(vertices and kernels) Green's functions. It is a $highly \
nonlinear$, strongly coupled system of four-dimensional integral
equations for the above-mentioned quantities. The kernels of
these integral equations are determined by the infinite series of
the corresponding skeleton diagrams [10-12]. It is a general
feature of $nonlinear$ systems that the number of exact solutions
(if any) $can \ not \ be \ fixed \ a \ priori$. Thus formally it
may have several exact solutions. These equations should be
also complemented by the corresponding Slavnov-Taylor (ST)
identities [10-12] which in general relate the above mentioned
lower and higher Green's functions to each other. These
identities are consequences of the exact gauge invariance and
therefore $"are \ exact \ constraints \ on \ any \ solution \ to
\ QCD"$ [10]. Precisely this system of equations can serve as an
adequate and effective tool for the nonperturbative approach to
QCD [13,14].

Among the above-mentioned Green's functions, the two-point Green's function
describing the full gluon propagator
\begin{equation}
iD_{\mu\nu}(q) = \left\{ T_{\mu\nu}(q)d(-q^2, \xi) + \xi L_{\mu\nu}(q) \right\} {1 \over q^2 },
\end{equation}
has a central place [10-15]. Here $\xi$  is a gauge fixing
parameter ($\xi = 0$,  Landau gauge) and $T_{\mu\nu}(q) =
g_{\mu\nu} - q_\mu q_\nu / q^2 = g_{\mu\nu } - L_{\mu\nu}(q)$.
Evidently, its free perturbative (tree level) counterpart is
obtained by simply setting the full gluon form factor $d(-q^2,
\xi)=1$ in Eq. (1.1). In particular, the solutions of the
above-mentioned SD equation for the full gluon propagator (1.1),
are supposed to reflect the complexity of the quantum structure
of the QCD ground state. As emphasized above, it is a $highly \
nonlinear$ system of four-dimensional integrals containing many
different, unknown in general, propagators, vertices and kernels
[10-12]. Because of truncation schemes this system becomes the
equation for the full gluon propagator only, but it remains
$nonlinear$, nevertheless. Different truncations could lead to
qualitatively different solutions, and the number of these solutions
may be increased only. Moreover, to clearly
distinguish between the exact or approximate solutions (if any), we
do not know even the complete set of boundary conditions 
to attempt to uniquely fix solution of the truncated equation. We
certainly know the boundary condition in the ultraviolet (UV)
limit because of asymptotic freedom and certainly we do not know
the corresponding boundary condition in the infrared (IR) because
precisely of confinement (at this stage it is not even clear
whether the two boundary conditions (in the UV and in the IR (if
it can be established)) will be sufficient to completely fix the
theory or not). Because of the above-discussed highly complicated
mathematical structure of the SD equation for the full gluon
propagator, there is no hope for exact solution(s). However, in
any case the solutions of this equation can be distinguished by
their behaviour in the deep IR limit (the UV limit is uniquely determined by   
asymptotic freedom), describing thus many
different types of quantum excitations and fluctuations of gluon
field configurations in the QCD vacuum. Evidently, not all of
them reflect the real structure of the QCD vacuum.

  The deep IR asymptotics of the full gluon propagator can be generally classified into the
three different types: 1) the IR enhanced (IRE) or IR singular
(IRS), 2) the IR finite (IRF) and 3) the IR vanishing (IRV) ones
(for references see the corresponding sections below). Let us
emphasize that any deviation in the behaviour of the full
gluon propagator in the deep IR domain from the free perturbative one
automatically assumes its dependence on a scale parameter (at
least one) in general different from QCD asymptotic scale
parameter, $\Lambda_{QCD}$. It can be considered as responsible
for the nonperturbative dynamics (in the IR region) in the QCD
vacuum models under consideration. If QCD itself is a confining theory, then   
such characteristic scale is very likely to exist. In what follows,
let us denote it, say $\Lambda_{NP}$. This is very similar to
asymptotic freedom which requires the above-mentioned asymptotic
scale parameter associated with nontrivial perturbative dynamics
in the UV region (scale violation). However, for calculation of the
truly nonperturbative VED we need not exactly the deep IR
asymptotics of the full gluon propagator, but rather its truly
nonperturbative part, which vanishes when the above-mentioned
nonperturbative scale parameter goes formally to zero, i.e., when only
the perturbative phase survives. So we define the truly
nonperturbative part of the full gluon form factor in Eq. (1.1)
as follows:

\begin{equation}
d^{NP}(-q^2, \Lambda_{NP}) = d(-q^2, \Lambda_{NP}) - d(-q^2, \Lambda_{NP} =0),
\end{equation}
which, on one hand, uniquely determines the truly nonperturbative
part of the full gluon propagator. On the other hand, the
definition (1.2) explains the difference between the truly
nonperturbative part $d^{NP}(-q^2)$ and the full gluon propagator
$d(-q^2)$ which is nonperturbative itself. Let us note in advance,
that in the realistic models for the full gluon propagator, the
limit $\Lambda_{NP} \rightarrow 0$ is usually equivalent to the
limit $-q^2 \rightarrow \infty$. In some cases, the model gluon
propagator does not depend explicitly on the nonperturbative
scale parameter (the dependence is hidden) then its behaviour at
infinity should be subtracted. In the realistic models  of
the full gluon propagator its truly nonperturbative part usually
coincides with its deep IR asymptotics underlying thus the strong
intrinsic influence of the IR properties of the theory on its
nonperturbative dynamics.

It is well known, however, that VED in general is badly divergent in quantum field theory, in particularly QCD [16]. Thus the main problem
is how to extract the truly nonperturbative VED which is relevant for the
QCD vacuum quantum model under consideration. It should be finite, negative and
it should have no imaginary part (stable vacuum).
Why is it so important to calculate it from first principles? i.e., on the basis of some realistic Ansatz for the full gluon propagator only. As was emphasized above, this quantity is important in its own right as being nothing else but 
the bag
constant (the so-called bag pressure) apart from the sign, by definition [16].
Through the trace anomaly relation [17] it helps in the correct estimation of  
such an important
phenomenological nonperturbative parameter as the gluon condensate introduced
in the QCD sum rules approach to resonance physics [18]. Furthermore, YM VED
assists
in the resolution of the $U(1)$ problem [19] via the Witten-Veneziano (WV) formula for
the mass of the $\eta'$ meson [20]. The problem is that the topological susceptibility [19-23]
needed for this purpose is determined by a two point correlation function from 
which perturbative contribution
is already subtracted, by definition [20,23-25]. The same is valid for the above-mentioned bag
 constant which is much more general quantity than the string tension because  
it is relevant
 for light quarks as well. Thus to calculate correctly the truly nonperturbative VED means to understand correctly the
structure of the QCD vacuum in different models.

We have already formulated a general method of calculation of the truly nonperturbative YM VED
in the axial gauge QCD in Ref. [26], where the Abelian Higgs model [27] of the
dual QCD [28]
 ground state was investigated. Moreover, we have calculated the truly
 nonperturbative VED (using particular method) in the covariant gauge QCD quantum vacuum model as well [29,30]. The
main purpose of this paper (section II) is to formulate precisely a general method of calculation of the truly nonperturbative quantum part of YM VED in the covariant gauge QCD.
In sections III, IV and V it is illustrated by
by considering different covariant gauge QCD quantum models of its ground state
by choosing three different types of the deep IR asymptotics of the full gluon
propagator, IRE, IRF and IRV, respectively. The conclusions
are presented in section VI.

\section{The truly nonperturbative vacuum energy density }

 In this section we formulate a general method of numerical calculation of the 
quantum part of the truly nonperturbative YM VED in the covariant gauge QCD.
Let us start from the gluon part of VED
which to-leading order (log-loop level $\sim \hbar$)\footnote{Next-to-leading and higher contributions
 (two and more vacuum loops) are numerically suppressed
by one order of magnitude in powers of $\hbar$ at least and are left for consideration elsewhere.}
is given by the effective potential for composite operators [7] as follows:
\begin{equation}
V(D) =  { i \over 2} \int {d^nq \over {(2\pi)^n}}
 Tr\{ \ln (D_0^{-1}D) - (D_0^{-1}D) + 1 \},
\end{equation}
where $D(q)$ is the full gluon propagator (1.1) and $D_0(q)$ is its
free perturbative (tree level) counterpart. Here and below the traces over
space-time and color group indices are understood.
The effective potential is normalized as $V(D_0) = 0$, i.e., the free perturbative vacuum is normalized to zero. In order to evaluate the
effective potential (2.1) we use the well-known expression
\begin{equation}
Tr \ln (D_0^{-1}D) = 8 \times \ln det (D_0^{-1}D) =
8 \times 4 \ln \left[ {3 \over 4 }d(-q^2) + {1 \over 4 } \right].
\end{equation}
It becomes zero (in accordance with the above mentioned normalization condition) when the full gluon form factor is replaced by its free perturbative counterpart. This composition does not depend explicitly on a gauge choice.
 Going over to four-dimensional ($n=4$) Euclidean space in Eq. (2.1), on account of (2.2), and evaluating some numerical factors, one obtains ($\epsilon_g = V(D)$)

\begin{equation}
\epsilon_g = - {1 \over \pi^2} \int dq^2 \ q^2 \left[ \ln [1 + 3 d(q^2)] -
{3 \over 4}d(q^2) + a \right],
\end{equation}
where constant $a = (3/4) - 2 \ln 2 = - 0.6363$ and the integration
from zero to infinity is assumed. Substituting the
definition (1.2) into the Eq. (2.3) and doing some trivial
rearrangement, one obtains

\begin{equation}
\epsilon_g = - {1 \over \pi^2} \int dq^2 \ q^2 \left[ \ln [1 + 3 d^{NP}(q^2, \Lambda_{NP})]
- {3 \over 4}d^{NP}(q^2, \Lambda_{NP}) \right] - {1 \over \pi^2} I_{PT},
\end{equation}
where we introduce the following notation

\begin{equation}
I_{PT} = \int dq^2 \ q^2 \left[ \ln [1 + {3d(q^2, \Lambda_{NP}=0) \over 1 + 3 d^{NP}(q^2, \Lambda_{NP})}] -
{3 \over 4}d(q^2, \Lambda_{NP}=0) + a \right].
\end{equation}
It contains the contribution which is mainly determined by the
perturbative part of the full gluon propagator, $d(q^2,
\Lambda_{NP}=0)$. The constant $a$ also should be included since
it comes from the normalization of the free perturbative vacuum
to zero. If we separate the deep IR region from the
perturbative one (which consists of the intermediate (IM) and UV
regions since the IM region remains $terra \ incognita$ in QCD) by
introducing the so-called soft cutoff explicitly then we get

\begin{equation}
\epsilon_g = - {1 \over \pi^2} \int_0^{q^2_0} dq^2 \ q^2 \left[ \ln [1 + 3 d^{NP}(q^2, \Lambda_{NP})]
 - {3 \over 4}d^{NP}(q^2,\Lambda_{NP}) \right] - {1 \over \pi^2}( \tilde{I}_{PT} + I_{PT}),
\end{equation}
where evidently

\begin{equation}
\tilde{I}_{PT} = \int_{q^2_0}^{\infty} dq^2 \ q^2 \left[ \ln [1 + 3 d^{NP}(q^2,\Lambda_{NP})]
 - {3 \over 4}d^{NP}(q^2,\Lambda_{NP}) \right].
\end{equation}
Thus the first integral represents contribution to YM VED which
is determined by the truly nonperturbative piece of the full
gluon propagator integrated over the deep IR region. In other
words, just this term is the truly nonperturbative contribution
to YM VED. This means that the two remaining terms in Eq. (2.6)
should be subtracted by introducing corresponding counter terms
into the effective potential. Thus in general the integral (2.5)
determining the contribution from the perturbative part of the
full gluon propagator and the integral (2.7) determining the
contribution from the perturbative region (IM plus UV) are of no
importance for our present consideration. The above-mentioned
necessary subtractions can be done in more sophisticated way by
means of ghost degrees of freedom (see below).

The effective potential at the log-loop level for the ghost degrees of freedom
is
\begin{equation}
V(G) = - i \int {d^np \over {(2\pi)^n}}
Tr\{ \ln (G_0^{-1} G) - (G_0^{-1} G) + 1 \},
\end{equation}
where $ G(p)$ is the full ghost propagator and $G_0(p)$ is its
free perturbative (tree level) counterpart. The effective
potential $V(G)$ is normalized as $V(G_0) = 0$. Evaluating
formally the ghost term $\epsilon_{gh} = V(G)$ in Eq. (2.8),
we obtain $\epsilon_{gh} = \pi^{-2}I_{gh}$. The 
integral $I_{gh}$ depends on the ghost
propagator, which remains arbitrary (unknown) within our
approach. In principle, we have to sum up all contributions 
to obtain the total VED (the confining quark part of the vacuum
energy density is not considered here). However, upon the
substitution of definition (1.2) into the integral over the
whole momentum range from zero to infinity (2.3), some terms appear
there which may have unphysical singularities below the
scale $\Lambda_{QCD}$ (integral (2.5)). Thus the initial VED
(2.3) is formal one, it suffers from unphysical singularities
briefly mentioned above and it is badly divergent as well. In
order to get physically
meaningful expression, one have to subtract two integrals (2.5) and (2.7)
from Eq. (2.3). We have done this subtraction with the help of
ghost term by imposing the following condition $\Delta =
\tilde{I}_{PT} + I_{PT} - I_{gh} = 0$. The
nonperturbative gluon contribution to VED is determined by subtracting unwanted
terms by means of the ghost contribution, i.e., defining
$\epsilon_g + \epsilon_{gh} = \epsilon_{YM}$ at $\Delta=0$. Thus
the truly nonperturbative YM VED becomes

\begin{equation}
\epsilon_{YM} = {1 \over \pi^2} \int_0^{q^2_0} dq^2 \ q^2 \left[ {3 \over 4}d^{NP}(q^2, \Lambda_{NP}) -
\ln [1 + 3 d^{NP}(q^2, \Lambda_{NP})]\right].
\end{equation}
In many
cases this subtraction is sufficient to obtain the expression  for the
truly nonperturbative YM VED. However, in some other cases the truly nonperturbative part of the full gluon propagator which enters Eq. (2.9) continues to suffer from unphysical singularities below the scale $\Lambda_{QCD}$ (see discussion at the end of section V). As it was noticed, some additional terms should be
included in our subtraction scheme in this case indicating that the chosen Ansatz for the full gluon propagator itself was not realistic.

A few general remarks are in order. In QCD nothing should
explicitly depend on ghosts. By contributing into the closed
loops only, the main purpose of their introduction is to cancel
unphysical degrees of freedom of gauge bosons (maintaining thus
the unitarity of $S$-matrix), for example to exclude the longitudinal
components, the above mentioned unphysical singularities below
the QCD scale, etc. This is the main reason why they are to be
considered together with gluons always. In
nonperturbative QCD in general and in our approach in particular
the ghost propagator (or equivalently the ghost self-energy) still
remains unknown (in this sense arbitrary) since the exact
ghost-gluon vertex (which enters the corresponding SD equation)
is not exactly known (in Refs. [31,32] some very specific
truncation scheme is used in order to derive particular
expression for this vertex). We know however, that the  ghost
propagator contribution to VED, regular or singular, should be combined with   
gluon contribution in order
to cancel exactly the above-mentioned unphysical singularities of the
gauge bosons which are inevitably present in any Ansatz for the
full gluon propagator. In other words, if one knows the ghost
propagator exactly then the above-mentioned cancellation  should
proceed automatically (as usual in perturbative calculus if, of
course, all calculations were correct). But if it is not known
exactly (as usual in nonperturbative calculus) then one has to
impose condition of cancellation as it was done in our case,
$\Delta=0$. Obviously, the above-mentioned condition of
cancellation was imposed in the most general form. Instead of the
introduction of some counter terms into the initial effective potential to cancel the most dangerous UV divergences
presented in the integral (2.5), we have used the ghost term for this
purpose as well. Thus our subtraction scheme is in
agreement with general physical interpretation of ghosts to
cancel all unphysical degrees of freedom of the gauge bosons
[10,33].

The expression (2.9) is our definition of the truly nonperturbative YM
VED as integrated out the truly nonperturbative part of the full gluon propagator over the deep IR region (soft momentum region, $0 \leq q^2
\leq q^2_0$). The soft cutoff $q_0^2$ (as a function of the nonperturbative scale) can be determined by the corresponding minimization procedure (see below).

\subsection{}

From this point it is convenient to factorize scale dependence of the truly
nonperturbative YM VED (2.9). As was already emphasized above, $d^{NP}(q^2)$ always contains at least one scale parameter ($\Lambda_{NP}$) responsible for
the nonperturbative dynamics in the model under consideration.
It is considered as free one within our general method, i.e.,
"running" (when it formally goes
to zero then the perturbative phase only survives in the model).
Its numerical value (if any) will be used at final stage only to evaluate numerically the corresponding truly nonperturbative YM VED (if any).
We can introduce dimensionless variables and parameters by
using completely extra scale (which is always fixed in comparison with $\Lambda_{NP}$), for example flavorless QCD asymptotic scale parameter $\Lambda_{YM}$
as follows:

\begin{equation}
  z = {q^2 \over \Lambda_{YM}^2}, \quad z_0 = {q_0^2 \over \Lambda_{YM}^2},
\quad b= {\Lambda^2_{NP} \over \Lambda^2_{YM}}.
\end{equation}
Here $z_0$ is a corresponding dimensionless soft cutoff while the parameter
$b$ has a very clear physical
meaning. It measures the ratio between nonperturbative dynamics, symbolized by
$\Lambda^2_{NP}$ and nontrivial perturbative dynamics (violation of scale, asymptotic freedom)
 symbolized by $\Lambda^2_{YM}$. When it is zero only perturbative
phase remains in the model. In this case, the gluon
form factor obviously becomes a function of $z$ and $b$, i.e.,  $d^{NP}(q^2)=d^{NP}(z, b)$
and the truly nonperturbative VED (2.9) is
($\epsilon_{YM} \equiv \epsilon_{YM}(z_0, b)$)

\begin{equation}
\Omega_g (z_0, b) = { 1 \over \Lambda_{YM}^4} \epsilon_{YM}(z_0, b),
\end{equation}
where the gluon effective potential
at a fixed scale, $\Lambda_{YM}$, [26,29,34] is introduced

\begin{equation}
\Omega_g \equiv \Omega_g (z_0, b) = {1 \over \pi^2}
\int_0^{z_0} dz \ z \left[ {3 \over 4}d^{NP}(z,b)   -  \ln [1 + 3 d^{NP}(z,b)] \right].
\end{equation}
This expression precisely allows us to investigate the dynamical structure of the YM vacuum. It is free of scale dependence since it has been already factorized in Eq. (2.11).
It depends only on $z_0$ and $b$ and a minimization procedure 
with respect to $b$,
$ \partial \Omega_g (z_0, b) / \partial b = 0$, (usually after integrated out in Eq. (2.12)) can provide a 
self-consistent relation between $z_0$ and $b$, that is we get 
$q_0$ as a function of $\Lambda_{NP}$.
 Let us note in advance that the final numerical results will depend on $\Lambda_{NP}$ only
 as it should be for the nonperturbative part of YM VED (see sections
III and IV below). Obviously, the minimization with respect to $z_0$ leads to trivial zero.
In principle, through the relation $\Lambda_{YM}^4 = q_0^4 z_0^{-2}$, it is possible to fix
soft cutoff $q_0$ itself, but this is not the case indeed since then $z_0$ can
not be varied.

\subsection{}

  On the other hand, the scale dependence can be factorized as follows:
\begin{equation}
  z={q^2 \over \Lambda^2_{NP}}, \quad  z_0={q_0^2 \over \Lambda^2_{NP}},
\end{equation}
i.e., $b=1$. For simplicity (but not loosing generality) we use the same notations
for the dimensionless set of variables and parameters as in Eq. (2.10). In
this case, the gluon
form factor obviously becomes a function of $z$ only, $d^{NP}(q^2)=d^{NP}(z)$
and the truly nonperturbative YM VED (2.9) becomes
\begin{equation}
\epsilon_{YM} (z_0) = {1 \over \pi^2} q_0^4 z_0^{-2}
\int_0^{z_0} dz \ z \left[ {3 \over 4}d^{NP}(z) - \ln [1 + 3 d^{NP}(z)]
\right].
\end{equation}
Evidently, to fix the scale is possible in the two different ways. In
principle, we can fix $\Lambda_{NP}$ itself, i.e., introducing
\begin{equation}
\tilde{\Omega}_g (z_0) = { 1 \over \Lambda^4_{NP}} \epsilon_{YM}(z_0)= {1 \over \pi^2}
\int_0^{z_0} dz \ z \left[ {3 \over 4}d^{NP}(z) - \ln [1 + 3 d^{NP}(z)] \right].
\end{equation}
However, the minimization procedure again leads to the trivial zero, which shows that
this scale can not be fixed.

In contrast to the previous case, let us fix the soft cutoff itself, i.e.,
setting [26,29,30]

\begin{equation}
\bar \Omega_g (z_0) = { 1 \over q_0^4} \epsilon_{YM}(z_0)= {1 \over \pi^2}  z_0^{-2}  \int_0^{z_0} dz \ z \left[ {3 \over 4}d^{NP}(z) -
\ln [1 + 3 d^{NP}(z)] \right].
\end{equation}
In this case the perturbative phase is recovered in the $z_0 \rightarrow \infty$ ($\Lambda_{NP} \rightarrow 0$) limit.
Now the minimization procedure with respect to $z_0$ is nontrivial. Indeed,
$ \partial \bar \Omega_g (z_0) / \partial z_0 = 0$, yields the following "stationary" condition

\begin{equation}
\int_0^{z_0} dz \ z \left[ {3 \over 4}d^{NP}(z) - \ln [1 + 3 d^{NP}(z)] \right] =
{1 \over 2} z_0^2 \left[ {3 \over 4}d^{NP}(z_0)   - \ln [1 + 3 d^{NP}(z_0)] \right],
\end{equation}
the solutions of which (if any) allow one to find $q_0$ as a function of $\Lambda_{NP}$.
On account of this "stationary" condition, the effective potential (2.16) itself becomes simpler for numerical calculations, namely
\begin{equation}
\bar \Omega_g (z_0^{st}) = {1 \over 2 \pi^2} \left[ {3 \over 4}d^{NP}(z_0^{st})
- \ln [1 + 3 d^{NP}(z_0^{st})] \right],
\end{equation}
where $z_0^{st}$ is a solution (if any) of the "stationary" condition (2.17) and
corresponds to the minimum(s) (if any) of the effective potential (2.16).
In the next sections, we illustrate how this method works by considering some  
quantum models of the covariant gauge QCD ground state explicitly.

\section{The IRE gluon propagator. ZME quantum model}

Today there are no doubts left that the dynamical mechanisms
of the important non-perturbative quantum phenomena such as quark confinement and
dynamical (or equivalently spontaneous) chiral symmetry breaking (DCSB) are closely
related to the complicated topologically nontrivial structure of
the QCD vacuum [1-4,10]. On the other hand, it also becomes clear that the nonperturbative
IR dynamical singularities, closely related to the nontrivial vacuum
structure, play an important role in the large distance behaviour
of QCD [35,36]. For this reason, any correct nonperturbative model of quark confinement
and DCSB necessarily turns out to be a model of the true QCD vacuum
and the other way around.

Our model of the true QCD ground state is based on the existence and importance
of such kind of the nonperturbative, quantum excitations
of the gluon field configurations (due to self-interaction of massless gluons only,
i.e., without explicit involving some extra degrees of freedom) which can
be effectively correctly described
by the $q^{-4}$ behaviour of the full gluon propagator in the deep IR domain   
(at small $q^2$) [29,30]. These excitations are topologically nontrivial also
since they lead to the nontrivial YM
VED (see below). Thus our main definition (1.2) becomes

\begin{equation}
d^{NP}(-q^2, \Lambda_{NP}) = d(-q^2, \Lambda_{NP}) - d(-q^2, \Lambda_{NP}=0)=
{\Lambda^2_{NP} \over (-q^2)}.
\end{equation}
In the above-mentioned papers [29,30] the nonperturbative scale was denoted as
$\bar \mu$, i.e., $\bar \mu \equiv \Lambda_{NP}$.
In this way we obtain the generally accepted form of the deep IR singular asymptotics for the full gluon propagator (for some references see below)

\begin{equation}
D_{\mu\nu}(q) \sim (q^2)^{-2} , \qquad q^2 \rightarrow 0,
\end{equation}
which may be refered equivalently to as the strong coupling regime [10].
It describes the zero momentum modes enhancement (ZMME) dynamical effect in QCD
at large distances. We prefer to use
simply ZME (zero modes enhancement) since we work always in momentum space.    
This is our primary dynamical assumption in this section. The main problem
due to this strong singularity is its correct treatment by the dimensional
regularization method [37] within the distribution theory [38], which 
was one of highlights of our previous publications [29,30] (see also Ref. [39]). There exist many arguments in favor of this behaviour:

a). Such singular behaviour of the full gluon propagator in
the IR domain leads to the area law for static quarks (indicative of confinement) within the Wilson loop approach [40].

b). The cluster property of the Wightman functions in QCD fails and this allows
such singular behaviour like (3.2) for the full gluon propagator in the deep IR
domain [41].

c). After the pioneering papers of Mandelstam
in the covariant (Landau) gauge [42] and Baker, Ball and Zachariasen in the axial gauge [43],
the consistency of the singular asymptotics (3.2) with direct solution of the  
SD equation for the full gluon propagator in the IR domain was repeatedly confirmed (see for example Refs. [13,14,44,45] and references therein).

d). Moreover, let us underline that
without this component in the decomposition of the full gluon propagator in continuum theory
it is impossible to "see" linearly rising potential between heavy
quarks by lattice QCD simulations [46] not involving
some extra (besides gluons and quarks) degrees of freedom.
This should be considered as a strong lattice evidence (though not direct) of the existence
and importance of $q^{-4}$-type excitations of gluon field configurations in the QCD vacuum.
There exists also direct lattice evidence that the zero modes are enhanced in the full gluon propagator indeed [47].

e). Within the distribution theory [38] the structure of the
nonperturbative IR singularities in four-dimensional Euclidean
QCD is the same as in two-dimensional QCD, which confines quarks
at least in the large $N_c$ limit [48]. In this connection, let
us note that  $q^{-4}$ IR singularity is the simplest
nonperturbative power singularity in four-dimensional QCD as well
as $q^{-2}$ IR singularity is the simplest nonperturbative power
singularity in two-dimensional QCD. The QCD vacuum is much more
complicated medium than its two-dimensional model, nevertheless,
the above-mentioned analogy is promising even in the case of the
nonperturbative dynamics of light quarks.

f). Some classical models of the QCD vacuum also invoke $q^{-4}$ behaviour of  
the gluon fields in the IR domain. For example, it appears in the QCD vacuum as
a condensation
of the color-magnetic monopoles (QCD vacuum is a chromomagnetic superconductor)
proposed by Nambu, Mandelstam and 't Hooft and developed by Nair and Rosenzweig
(see Ref. [49] and references therein. For recent developments in this model
see Di Giacomo's contribution in Ref. [1]) as well as in the classical mechanism of the confining medium [50] and in effective theory for the QCD vacuum proposed in Ref. [51].

g). It is also required to derive the heavy quark potential within
the recently proposed exact renormalization group flow equations approach [52].

h). It has been shown in our papers that the singular behaviour
(3.2) is related directly to light quarks confinement and DCSB
[29,30]. Moreover, a very good agreement has been obtained  with the
phenomenological values of the topological susceptibility, the
mass of the $\eta'$ meson and the gluon condensate [21,22].

Thus we consider our main Ansatz (3.1), (3.2) as physically
well-motivated. Let us emphasize that $d^{NP}(-q^2, \xi) =
\Lambda^2_{NP} / (-q^2)$ is the truly nonperturbative part of the
full gluon propagator since it vanishes in the perturbative limit
($\Lambda^2_{NP} \longrightarrow 0$, when the perturbative phase
survives only) and simultaneously it correctly reproduces the
deep IR asymptotics of the full gluon propagator, i.e.,
$d^{NP}(-q^2)$ coincides with $d^{IR}(-q^2)$.

\subsection{}

The truly nonperturbative YM VED is given now by Eq. (2.9) with
$d^{NP}(q^2) = \Lambda^2_{NP} / q^2$ in Euclidean space.
Let us first introduce the A-type set of dimensionless variables (2.10). Then
$d^{NP}(q^2)$ becomes
$d^{NP}(z,b) = b / z$. Performing almost trivial integration in the effective potential
at a fixed scale (2.12), one obtains
\begin{equation}
\Omega_g (z_0, b) = {1 \over 2 \pi^2}
\left[ 9 b^2 \ln \left(1 + {z_0 \over 3 b} \right)  - {3 z_0 \over 2} b -
z_0^2 \ln \left(1 + {3 b \over z_0} \right) \right].
\end{equation}
It is easy to show that as a function of $b$, the effective potential (3.3) linearly approaches
zero from below and it diverges also linearly at infinity while as a function of $z_0$ itself it approaches zero from above and also diverges as
$\sim -z_0$ at infinity. Thus as a function of $b$ it has a local minimum
(relating $b$ to $z_0$) at which the truly nonperturbative YM VED will be always finite and negative. The minimization procedure with respect to $b$,
 $ \partial \bar \Omega_g (z_0; b) / \partial b = 0$, yields the following
 "stationary" condition, $\nu = 4 \ln (1 + (\nu / 3))$,
where $\nu = z_0 / b$. Its solution is $\nu^{min} = 2.2$. Using this "stationary" condition, the effective potential (3.3) can be written down as follows:

\begin{equation}
\Omega_g (\nu^{min}, b) = {b^2 \nu^{min} \over 2 \pi^2}
\left[ {3 \over 4} -  \nu^{min} \ln \left(1 + {3 \over \nu^{min}} \right) \right] =
 - 0.1273 b^2,
\end{equation}
so the truly nonperturbative YM VED (2.11) becomes
\begin{equation}
\epsilon_{YM} = - 0.1273 \times \Lambda^4_{NP},
\end{equation}
where the relation $\Lambda^4_{NP} = b^2 \Lambda^4_{YM}$ has been
already used. Determined in this way, it is always finite (since
characteristic scale of our model $\Lambda_{NP}$ is finite,
evidently it can not be arbitrary large), automatically negative
(as it should be for the truly nonperturbative energy) and it has
no imaginary part (stable vacuum). Obviously the characteristic
scale of our model $\Lambda_{NP}$ can not be determined within
the YM theory alone. Its numerical value should be taken from
good physical observable in full QCD by implementing the
physically well-motivated scale-setting scheme. Precisely this
has been done in our papers [29,30] where the nonperturbative VED
was numerically evaluated from first principles.  Moreover, in
recent publications [21,22] it is shown that our numerical
results are of the necessary order of magnitude in order to nicely
saturate the large mass of $\eta'$ meson in the chiral limit as
well as the phenomenological value of the topological
susceptibility. Thus the existence of the nontrivial VED in ZME
quantum model, which agrees well with QCD topology, is one more
serious argument in its favor. It is worthwhile to present
numerical value for the soft cutoff in terms of $\Lambda_{NP}$,
namely $q_0 = 1.48324 \Lambda_{NP}$.
 It follows from the solution of the "stationary" condition, of course.

\subsection{ }

It is instructive to calculate the truly nonperturbative YM VED by
choosing the B-type set of dimensionless variables (2.13).
Then $d^{NP}(q^2) = \Lambda^2_{NP} / q^2$ becomes
$d^{NP}(z) = 1 / z$. Performing almost trivial integration in the effective potential
at a fixed scale (2.16) in this case, one obtains
\begin{equation}
\bar \Omega_g (z_0) = {1 \over 2 \pi^2} z_0^{-2}
\left[ 9 \ln \left(1 + {z_0 \over 3 } \right)  - {3 \over 2} z_0 -
z_0^2 \ln \left(1 + {3 \over z_0} \right) \right].
\end{equation}
It is easy to show now that as a function of $z_0$, the effective potential (3.7)
diverges as $\sim z_0^{-1}$ at small $z_0$ and converges as
$\sim - z_0^{-1}$ at infinity (perturbative limit), see Fig. 1.
Thus as a function of $z_0$ it has a local minimum at
$z_0 = 4 \ln (1 + (z_0 / 3))$,
the so-called "stationary" condition in this case. Its solution again is
$z_0^{min} = 2.2$. At "stationary" state the effective potential (3.6)
can be written down as follows:
\begin{equation}
\bar \Omega_g (z_0^{min}) = { 1 \over 2 \pi^2}
\left[ {3 \over 4} (z_0^{min})^{-1} -  \ln \left(1 + {3 \over z_0^{min}}
\right) \right] = - 0.0263,
\end{equation}
so the truly nonperturbative YM VED (2.16) becomes
\begin{equation}
\epsilon_{YM} = - 0.0263 q_0^4 = - 0.1273 \times \Lambda^4_{NP},
\end{equation}
where the relation  $q_0^4 = (z_0^{min})^2 \Lambda^4_{NP}$ has been already used.
Thus we have explicitly demonstrated that truly nonperturbative YM VED does not
indeed depend on how one introduces dimensionless variables
into the effective potential, i.e., $\epsilon_{YM} = \Lambda^4_{NP} \Omega_g (\nu^{min}, b)
= q_0^4 \bar \Omega_g (z_0^{min})= - 0.1273 \Lambda^4_{NP}$. In some cases, the B-type
calculation is preferable. For example, to calculate confining quark contribution into
the total VED is much easier using precisely this set of the dimensionless variables
(see our papers [29,30] and next section as well).

\begin{figure}[h]
\vspace*{17pt}
\psfig{file=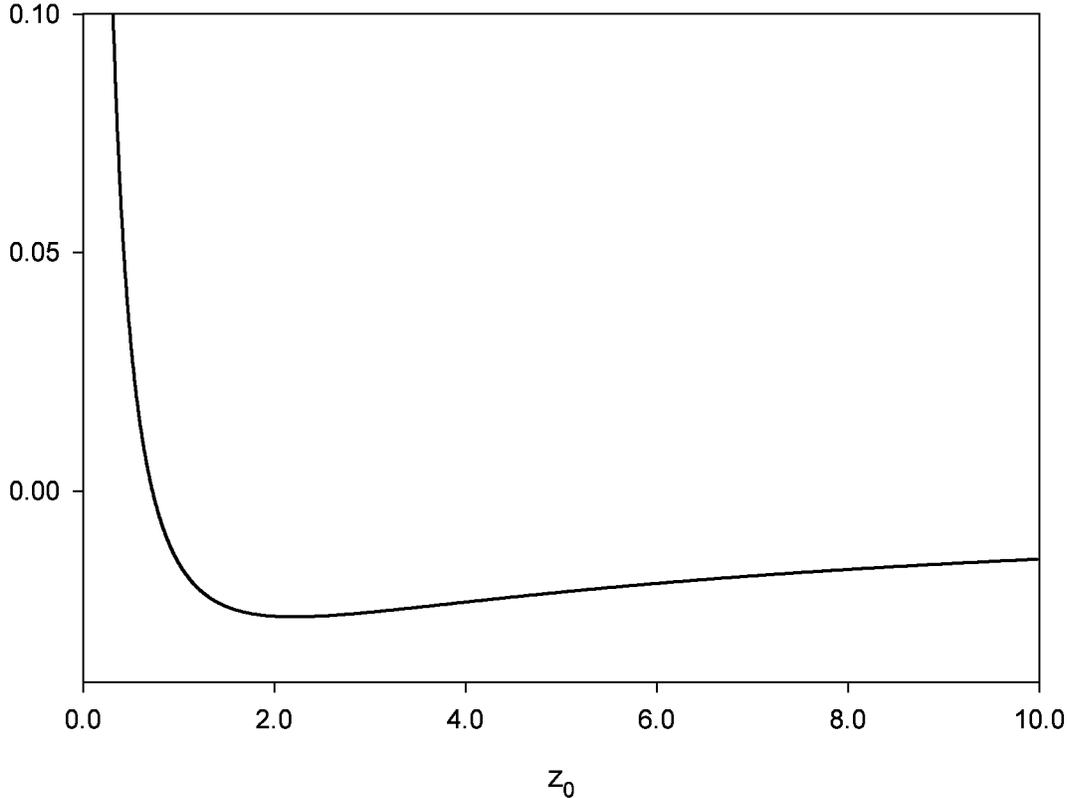}
\vspace*{13pt}
\caption{Effective potential (3.7) as a function of $z_0$.}
\end{figure}

\section{The IRF gluon propagator }

Let us consider now a possible IRF behaviour of the full gluon propagator (in  
the Landau gauge) in the
deep IR domain, which was suggested by recent lattice calculations
in Ref. [53]. The main definition (1.2) in this case becomes
\begin{equation}
d^{NP}(-q^2, M) = d (-q^2, M) - d (-q^2, M=0) = { Z A M^{2 \alpha}(- q^2)
\over ( - q^2 + M^2)^{1+ \alpha}}.
\end{equation}
Here $M$ is the mass scale parameter responsible for the nonperturbative dynamics in this model, i. e., $M= \Lambda_{NP}$ in our notation.
When the parameter $M$ formally goes to zero, the perturbative phase only remains in this model.
 Again as in previous case, the truly nonperturbative part vanishes
in the perturbative limit ($M \longrightarrow 0$) and it reproduces the IR asymptotics of the full gluon propagator correctly as well.
 The best estimates for the parameters $M$ and $A$ are $M = (1020 \pm 100 \pm 25) \ MeV$ and
  $A= (9.8 + 0.1 - 0.9)$
. As it was emphasized above, the numerical value of the parameter $M$ will be
used only at final stage in order to estimate numerically the truly nonperturbative YM VED
in this model. The exponent in general is $\alpha = 2 + \delta$, where
$\delta >0$ and small, while $Z \approx 1.2$ is the renormalization constant.

In this case, it is convenient to choose the B-type set of variables and parameters (2.13).
 Then $d^{NP}(q^2)$ in Euclidean space becomes
\begin{equation}
d^{NP}(z) = { a_1 z  \over (1+ z)^{1+ \alpha}},
\end{equation}
where the parameter $a_1 = Z A = 11.76$ is fixed. Substituting this into the
effective potential (2.16), one obtains
\begin{equation}
\bar \Omega_g (z_0; a_1) = { 1 \over  q_0^4} \epsilon_{YM} =
 - {1 \over \pi^2} \times z_0^{-2} \Bigl\{ I_1(z_0; a_1)- I_2(z_0; a_1) \Bigr\},
\end{equation}
where integrals are given as follows:

\begin{eqnarray}
I_1 (z_0; a_1) &=& \int \limits_0^{z_0} dz\, z\, \ln \Big( 1 +
{ 3a_1 z \over (1 + z)^{1+\alpha}} \Big),  \nonumber\\
I_2 (z_0; a_1) &=& { 3 a_1 \over 4} \int \limits_0^{z_0} dz\, z\,
{z \over (1 + z)^{1+\alpha}}.
\end{eqnarray}
The asymptotic behaviour of the effective potential (4.3) depends
on the asymptotic properties of the integral $I_1(z_0; a_1)$ since the integral
$I_2(z_0; a_1)$ in Eq. (4.4) can be taken explicitly, namely (in what follows
in this section, $\alpha =2$)

\begin{equation}
I_2 (z_0; a_1) = { 3 a_1 \over 4} \Big( \ln (1 + z_0)+
2 [ (1 + z_0)^{-1} - 1] - {1 \over 2} [ (1 + z_0)^{-2} - 1] \Big).
\end{equation}
From these expressions it is almost
obvious that asymptotics of the effective potential (4.3) at $z_0 \rightarrow 0, \infty$ to-leading order can be easily evaluated analytically.
The effective potential (4.3) as a function of the soft cutoff $z_0$ has
two local minimums (see below).
The corresponding "stationary" condition can be evaluated as follows:
\begin{equation}
[I_1(z_0; a_1) - I_2(z_0; a_1)] = {1 \over 2} z_0^2 \Bigl\{ \ln \Big( 1 + {3 a_1 z_0
\over (1+z_0)^3 } \Big) - {3a_1 z_0 \over 4 (1+z_0)^3} \Bigr\}.
\end{equation}
Using this "stationary" condition, the effective potential (4.3) at "stationary" state becomes
\begin{equation}
\bar \Omega_g (z_0^{st}; a_1) = - {1 \over 2 \pi^2} \Bigl\{ \ln \Big( 1 + {3 a_1 z_0^{st}
\over (1+z_0^{st})^3 } \Big) - {3a_1 z_0^{st} \over 4 (1+z_0^{st})^3} \Bigr\},
\end{equation}
where $z_0^{st}$ is a solution(s) to the "stationary" condition (4.6).
The two solutions of the "stationary" condition (4.6) corresponding to the two
local minimums are $z_0^{st} = 0.19$ and $z_0^{st} = 2.37$ with almost equal numerical
values for the corresponding effective potentials at "stationary" states, namely
$\bar \Omega_g (0.19; a_1)= - 0.0309$ and $\bar \Omega_g (2.37; a_1)=
- 0.0310$, respectively. However, the numerical values of the nonperturbative
YM VED (4.3) are drastically different,
\begin{equation}
\epsilon_{YM} (0.19) = - 0.0309 q_0^4(0.19) = - 0.00123 \times M^4
\end{equation}
and
\begin{equation}
\epsilon_{YM} (2.37) = - 0.0310 q_0^4(2.37) = - 0.174 \times M^4,
\end{equation}
where the relation  $q_0^4 = (z_0^{min})^2 M^4$
and the corresponding values of $z_0^{min}(\equiv z_0^{st})$ were applied. How 
to distinguish between the two solutions
for the truly nonperturbative YM VED (4.8) and (4.9)? This question is discussed in the following.


\subsection{Discussion}

In the first case, on account of the numerical value of the nonperturbative scale
$M \approx 1 \ GeV$, Eq. (4.8) numerically becomes
\begin{equation}
\epsilon_{YM} (0.19) = - 0.00123 \ GeV^4.
\end{equation}
It is the same order of magnitude as VED due to instantons [22]. Thus summing up this and
 instantons with ZME values, one obtains a fair
agreement with chiral QCD topology [20]. Also the soft cutoff in this case
is $q_0 \approx 0.463 M \approx 463 \ MeV$. This is quite reasonable value for the deep IR
region (in continuum theory) where the smooth-type behaviour of the full
gluon propagator effectively takes place.

In the second case, on account of the numerical value of the nonperturbative scale
$M \approx 1 \ GeV$, Eq. (4.9) numerically becomes
\begin{equation}
\epsilon_{YM} (2.37) = - 0.174 \ GeV^4.
\end{equation}
In Refs. [21,22]
an analytical formalism has been developed which allows one to calculate the topological
 susceptibility as a function of the truly nonperturbative YM VED. The
corresponding expression is
\begin{equation}
\chi_t = - \left( { 4 \xi \over 3} \right)^2 \epsilon_{YM},
\end{equation}
where parameter $\xi$ has two different values, namely $\xi_{NSVZ}= 2/11$ and
$\xi_{HZ}= 4/33$ (see Ref. [22]). Evaluating (4.12) numerically, on account of
(4.11), one obtains, $\chi_t^{NSVZ} = (550.8 \ MeV)^4$ and
$\chi_t^{HZ} = (259.6 \ MeV)^4$, while its phenomenological value is,
$\chi_t^{phen} = (180.36 \ MeV)^4$.
Thus, Eq. (4.11) substantially overestimates the phenomenological value of the
topological susceptibility (in both modes) and consequently the mass of $\eta'$
meson in the chiral limit, indeed. The soft cutoff in this case is $q_0 \approx
1.54 M \approx 1.54 \ GeV$.  It is also hard to imagine that the deep IR region
(in continuum theory) can be effectively extended up to $\approx 1.54 \ GeV$ especially for the smooth-type behaviour of the full gluon propagator there.
The continuum limit of the scale parameter $M$ is not known, so its realistic
numerical value still remains to be well-established, and so does the selection
from solutions, Eqs. (4.8) and (4.9). Let us note, that in accordance with the 
general scheme of our method we distinguish the nonperturbative scale of this model from the perturbative one but for simplicity we retain the same notation. 
Evidently, one will obtain the same numerical results for the truly nonperturbative YM VED by choosing the set
of variables of A type.

\section{The IRV gluon propagator}

The IRV full gluon propagator is represented by the so-called
Zwanziger-Stingle (ZS) formula [54,55]
\begin{equation}
d(-q^2) =  { (-q^2)^2 \over (-q^2)^2 + \mu^4 },
\end{equation}
in the whole range, where $\mu^4$ is again the mass scale
parameter responsible for the non-perturbative dynamics in this
model, i.e., $\mu \equiv \Lambda_{NP}$, in our notation. When it
is zero then the ZS gluon propagator (5.1) becomes free
perturbative one, indeed. Though the full gluon propagator (5.1)
is nonperturbative itself,
 however its truly
nonperturbative part is determined by the subtraction (1.2), i.e,
\begin{equation}
d^{NP}(q^2) = d(q^2, \mu^4) - d(q^2, \mu^4=0) = - { \mu^4 \over (-q^2)^2 + \mu^4 }.
\end{equation}
Since this expression is rather simple, it will be instructive to
perform calculations in both schemes, A (2.10) and B (2.13). So
let us start from A scheme.

\subsection{ }

Within the A-type set of variables (2.10), $d^{NP}(q^2)$ from Eq. (5.2) becomes
$d^{NP}(z, b) = - (b^2 / b^2+z^2)$ (Euclidean space). After the integration
over four-dimensional Euclidean space in Eq. (2.12), one obtains
\begin{eqnarray}
\Omega_g (z_0, b) = {1 \over 8 \pi^2} \Bigl\{ - 8b^2 \ln 2b^2 + 8 b^2 \ln (2b^2
- z_0^2) &+& (b^2 + 4 z_0^2) \ln (b^2 + z_0^2) \nonumber\\
&-& 4 z_0^2 \ln (z_0^2 - 2 b^2) -
b^2 \ln b^2 \Bigr\}.
\end{eqnarray}
From this expression it follows obviously that the effective
potential (5.3) at any finite relation between the soft cutoff
$z_0$ and parameter $b$ will always contain the imaginary part,
which is a direct manifestation of the vacuum instability [56] in
this model. Its asymptotics at $b \rightarrow 0, \infty$
to-leading order can be easily evaluated analytically. Omitting
all intermediate calculations, one finally obtains, $\Omega_g
(z_0, b) \sim_{b \rightarrow 0} - (9 / 8 \pi^2) b^2 \ln b^2$ and
$\Omega_g (z_0, b) \sim_{b \rightarrow \infty} - (1 / 8 \pi^2) [3
+4 \ln (-2)] z_0^2$, confirming the vacuum instability. Let us
also consider the corresponding formal "stationary" condition, $
\partial \Omega_g (z_0, b) / \partial b = 0$, which yields
\begin{equation}
3 t_0^2 + (1 + t_0^2) \ln (1 + t_0^2 ) + 8 (1 + t_0^2) \ln ( 1 - {t_0^2 \over 2}) = 0,
\end{equation}
where $t_0^2 = (z_0^2 / b^2)$. It has only trivial solution $t_0 = z_0=0$.

Thus the vacuum of this model is unstable, indeed, so it has no relation to quark
confinement and DCSB. Our conclusion is in full agreement
 with conclusion driven in Ref. [57].
The particular-type expressions for the dressed-quark-gluon vertex free
from ghost contributions were used in their investigation. Our result, however, is a
general one since it does not require the particular choice of the dressed-quark-gluon vertex.

\subsection{ }

Within the B-type set of variables (2.13), $d^{NP}(q^2)$ from Eq. (5.2) becomes
$d^{NP}(z) = - (1 / 1+z^2)$ (Euclidean space). After almost trivial integration
over four-dimensional Euclidean space in Eq. (2.16), one obtains
\begin{equation}
\bar \Omega_g (z_0) = {1 \over 8 \pi^2} z_0^{-2} \Bigl\{ -8 \ln 2 +
8 \ln (2 - z_0^2) + (1+ 4 z_0^2) \ln (1 + z_0^2) - 4 z_0^2 \ln (z_0^2 - 2) \Bigr\}.
\end{equation}
From this expression it obviously follows that the effective
potential at any finite value of the soft cutoff $z_0$ will
always contain the imaginary part, which is a direct
manifestation of the vacuum instability [56] as it was indicated
above. Its asymptotics at $z_0 \rightarrow 0, \infty$ to-leading
order can be easily evaluated analytically. Omitting all
intermediate calculations, one finally obtains, $\bar \Omega_g
(z_0) \sim_{z_0 \rightarrow 0} - (1 / 8 \pi^2) [3 + 4 \ln (-2)]$
and $\bar \Omega_g (z_0) \sim_{z_0 \rightarrow \infty} (9 / 8
\pi^2) z_0^{-2} \ln z_0^2$, so the vacuum of this model is
unstable, indeed. In order to confirm this, let us consider the
corresponding formal "stationary" condition which is
\begin{equation}
3z_0^2 + (1 + z_0^2) \ln (1 + z_0^2 ) + 8 (1 + z_0^2) \ln ( 1 - {z_0^2 \over 2}) = 0.
\end{equation}
It has only trivial solution $z_0=0$.

In Ref. [57] it was proposed the modification of the ZS
propagator (5.1) which took into the consideration the
renormgroup improvements to-leading
 order for the running coupling constant in the UV region, namely
\begin{equation}
d(-q^2) =  { (-q^2)^2 \over (-q^2)^2 + \mu^4 } \times {const \over \ln \left( \tau +
 { q^2 \over \Lambda^2_{QCD}} \right)}.
\end{equation}
Here $const$ obviously depends on the first coefficient of the $\beta$-function
and an unphysical parameter $\tau$ is introduced in order to regulate unphysical singularity -- Landau pole -- at $q^2=\Lambda^2_{QCD}$ (Euclidean space).    
The truly nonperturbative part now is

\begin{equation}
d^{NP}(q^2) = d(q^2, \mu^4) - d(q^2, \mu^4=0) = - { \mu^4 \over (-q^2)^2 + \mu^4 }
\times {const \over \ln \left( \tau + { q^2 \over \Lambda^2_{QCD}} \right)}.
\end{equation}
However, it is possible to show that YM VED continues to contain imaginary part
in this case as well. It is worth noting, that in derivation of the corresponding expression for YM VED (2.9) all terms depending in general on some unphysical parameters (in this case $\tau$) should be additionally subtracted by means of ghosts (as it was mentioned above in Section II just after Eq. (2.9)).
Concluding, let us note that neither (5.2) nor (5.8) coincides with deep IR asymptotics of the corresponding full gluon propagators (5.1) and (5.7).

\section{Conclusions}

In summary, we have formulated a general method how to numerically calculate
the quantum part of the truly nonperturbative YM VED (the bag constant, apart
from the sign, by definition) in the covariant gauge QCD quantum
models of its ground state using the effective potential approach for composite
operators. It is defined as integrated out the truly nonperturbative part of the full
gluon propagator over the deep IR region (soft momentum region), Eq.
(2.9). The nontrivial minimization procedure makes it possible to determine the
value of the soft cutoff as a function of the corresponding nonperturbative scale parameter
which is inevitably present in any nonperturbative
full gluon propagator model. If the chosen Ansatz for the full gluon propagator is realistic one,
then our general method gives the truly nonperturbative YM VED which
is always finite, automatically negative and it has no imaginary part (stable vacuum) (sections III and IV).
 Its numerical value does not, of course, depend on how one introduces the scale dependence by choosing different scale parameters as it was described above in subsections A and B of section II, i.e., both set of variables lead to the same numerical value of the truly nonperturbative YM VED.

From comparison of Eqs. (2.3) and (2.9), a prescription can be derived how one
can obtain the relevant expression for the truly nonperturbative YM VED.       
For this purpose the full gluon propagator in Eq. (2.3) should be replaced
by its truly nonperturbative part in accordance with Eq. (1.2).
The constant $a$ should be
omitted (it was already explained why) and the soft cutoff $q_0^2$ on the upper
limit should be introduced. Now it looks like the UV cutoff. Nevertheless let  
us underline
once more that it separates the deep IR region from the perturbative one, which includes
the IM region as well. It has a clear physical meaning as determining the range
where the deep IR asymptotics of the full gluon propagator is valid. By definition it can not be
arbitrary large as the UV cutoff is. As far as one chooses Ansatz for
 the full gluon propagator, the separation
"NP versus PT" is exact because of the definition (1.2).
The separation "soft versus hard"
 momenta is also exact because of the above-mentioned minimization procedure.
 Thus the proposed determination of the truly nonperturbative YM VED is uniquely defined.
The nontrivial minimization procedure can be done only by two
ways. First, to minimize the effective potential at a fixed scale
(2.11), (2.12) with respect to the physically meaningful
parameter. When it is zero, the perturbative phase only survives
in all models of the QCD ground state. Equivalently, we
can minimize the auxiliary effective potential (2.16) as a
function of the soft cutoff $z_0$ itself.                                      
When it goes to infinity then again the perturbative phase survives only. On the other hand, both effective potentials (2.12) and (2.16) should go to zero in 
the perturbative limit since the perturbative contributions have been already  
subtracted from the very beginning (see section 2).                          
As was underlined
above, both methods lead to the same numerical value for the
truly nonperturbative YM VED.

We have shown explicitly that the IRE gluon propagator (3.2) as
well as IRF (4.1) correspond to nontrivial VED which is always
finite, negative and it has no imaginary part (stable vacuum). In
this way they reflect some physical types of excitations of gluon
field configurations in the QCD vacuum. At the same time, the IRV
gluon propagators (5.1) and (5.2) lead to unstable vacuum and
therefore are physically impossible. However, these results are
by no means general. For example, to come to the same conclusion
for the IRV gluon propagator obtained and investigated in Refs.
[31,32] it is necessary to proceed along with lines of our
method. Thus the proposed method is precisely general one and
each particular model for the full gluon propagator should be
separately analyzed within its framework. However, it seems to us that the     
unstable vacuum is a fundamental defect of all vacuum models based on the      
IRV-type behaviour of the full gluon propagator. It is worthwhile also noting 
that, in contrast to IRE gluon propagator, the smooth behaviour of
the full gluon propagator in the IR domain is hard to relate to
quark confinement and DCSB.

Thus our method can serve as a test of any different QCD vacuum models (quantum
or classical) since it provides an exact
criterion for the separation ``stable versus unstable vacuum''.
Vacuum stability in classical models is important as well. For
example, we have already shown [26] that the vacuum of the
Abelian Higgs model without string contributions is unstable
against quantum corrections.

There is no general method of calculation of the confining
quark contribution to the total VED. In quantum theory it heavily depends
on the particular solutions of the corresponding quark SD equation,
on account of the chosen Ansatz for the full gluon propagator.
If it is correctly calculated then it is of opposite sign to the nonperturbative gluon part and it is one order of magnitude less (see, for example our papers [21,22,29,30]). Our method is not a solution for the fundamental
badly divergent problem of VED in QCD. Moreover, it is even not necessary
to deal with this problem. What is necessary indeed, is to be
able to extract the finite part of the truly nonperturbative VED in a self-consistent way. Just this is provided by our method which thus can be applied
to any nontrivial QCD vacuum quantum/classical models.

In conclusion, let us make some remarks. In some cases together
with the nonperturbative scale some other parameter(s) should be
considered as "running" in accordance with the general scheme of
our method. For example, such situation will arise in the IRF
model gluon propagator suggested by lattice calculations in Ref.
[58] (see also Ref. [59]). In this case the general procedure of
calculation of the truly nonperturbative YM VED (if any) remains,
of course, unchanged. However, due to some technical details
(for example, the corresponding "stationary" condition (2.17)
will be more complicated) this case requires
 a separate consideration. A brief recent reviews on
both continuum and lattice gluon propagators can be found in Refs. [15,53].
An attempt of VED calculation by introduction of rather controversial gluon    
mass was made in a recent paper [60].

\acknowledgments

One of the authors (V.G.) is grateful to M. Polikarpov, M. Chernodub, A. Ivanov
and especially to V.I. Zakharov for useful and critical discussions which led finally authors to the formulation of a general method presented here.
He also would like to thank H. Toki for many discussions on this subject during
his stay at RCNP, Osaka University. It is a special pleasure to thank for the  
referee's remarks, comments and suggestions which substantially improved the content of the present paper.  
This work was supported by the OTKA grants No: T016743 and T029440.

\vfill

\eject

\end{document}